# Spatiotemporal optical vortex (STOV) polariton

M. S. Le, S. W. Hancock, N. Tripathi, and H. M. Milchberg

*Institute for Research in Electronics and Applied Physics and Dept. of Physics, University of Maryland, College Park, Maryland 20742, USA*

We confirm the existence of a new bulk medium quasiparticle with transverse orbital angular momentum (tOAM) and elucidate its physical origin. The tOAM structure is driven by torques induced by the ponderomotive force of the light in the medium, originating from the magnetic Lorentz force, even for weak light fields. There are two contributions to the material tOAM: one part depends on the pulse spatial shape change as it passes through an interface, and the other part depends on dispersion. The results, from first principles particle-in-cell simulations of a simple plasma, are in excellent agreement with our theory for general dielectric media [S. W. Hancock *et al*., Phys. Rev. Lett. **127**, 193901 (2021)]. For plasma, there is agreement up to near-critical densities and near-relativistic field strengths.

In the past few years, a new form of structured light has been the subject of rapidly expanding research activity: spatiotemporal optical vortices (STOVs), polychromatic electromagnetic structures which carry orbital angular momentum (OAM) oriented transverse to propagation, with the associated phase winding and energy density circulation in a spatiotemporal plane [1,2]. This is in contrast with well-known monochromatic beams carrying longitudinal OAM [3], such as Laguerre-Gaussian modes, whose phase and energy density circulates in spatial degrees of freedom [3].

STOVs were first observed as emergent toroidal electromagnetic (EM) structures, akin to smoke rings, threaded by the propagation axis and initiated by the extreme spatiotemporal phase shear accumulated during arrested self-focusing collapse of intense femtosecond pulses in air [1]. STOVs mediate EM energy flow in intense propagating pulses over widely different intensity regimes, from nonrelativistic filamentation in Kerr media to relativistic filamentation in plasmas [1,4]. As they are carried by pulses, STOVs are necessarily polychromatic [5]. Their generation by spatiotemporal phase shear led to the realization that one could generate freely propagating STOVs by employing a $4f$ pulse shaper to first apply phase shear to a pulse in the spatiospectral domain and then return it to the spatiotemporal domain [6–8]. Other efforts followed [9] using this technique, with still others exploring the use of metasurfaces [10,11]. Further experiments demonstrating transverse OAM (tOAM) conservation in second harmonic generation provided confirmation that tOAM can be carried by photons [12,13].

In parallel with our experimental work, we developed a linear theory of mode structure and tOAM of STOVs in a general dispersive medium that predicted half-integer intrinsic tOAM and the existence of a quasiparticle we called a "STOV polariton", which embodies the tOAM-carrying material response [2]. The theory is in excellent agreement with measurements of freely propagating STOVs in air [2] and experiments measuring the tOAM transfer by spatiotemporal refractive index perturbations on general light pulses, a process we dubbed "spatiotemporal torquing of light" [14,15]. However, ref. [2] was based on a given constant material dispersion and contained no dynamical microscopic model for the response of the medium, and so the physics of how tOAM is delivered to matter was not understood.

In this paper, we confirm the existence of this new quasiparticle, the STOV polariton, and determine the value of its transverse orbital angular momentum. We develop a physical picture of the polariton as a tOAM structure driven by torques induced by the ponderomotive force of the light, mediated by the magnetic Lorentz force– remarkably, even for weak field strengths. In a first principles approach, we use high resolution particle-in-cell (PIC) simulations of a uniform density slab of fully ionized, collisionless hydrogen plasma, directly tracking the tOAM of the particles and fields. There are two advantages of this approach: (1) we use only Maxwell's equations and the Lorentz force, with no assumptions about constitutive or dispersion relations of a medium, and (2) the response can be explored from the linear through relativistic regimes.

Our linear theory is briefly reviewed in Appendix A. One solution of the theory's spatiotemporal paraxial wave equation (A1) is a STOV pulse with transverse orbital angular momentum. In vacuum or in media, the simplest STOV pulse with winding number or topological charge $l$ $(0, \pm1, \pm2, ..)$ has the following form near its waist for $z/z_{0x} << 1$ and $z/z_{0y} << 1$ [2],

$$A_{l,\alpha}(x, y, \xi : z = 0) = A^{(0)} \left( \frac{\xi}{w_{0\xi}} + \mathrm{sgn}(l) \frac{ix}{w_{0x}} \right)^{|l|} e^{-(x^2/w_{0x}^2 + y^2/w_{0y}^2)} e^{-\xi^2/w_{0\xi}^2} , \tag{1}$$

where the various parameters are defined in Appendix A. Here, we take both the field polarization and the tOAM to be directed along $y$ to eliminate the contribution of photon spin to the problem, $\{w_{0x}, w_{0y}, w_{0\xi}\}$ are scale widths of the pulse along $\{x, y, \xi\}$, $z_{0x} = \frac{1}{2}k_0 w_{0x}^2$ and $z_{0y} = \frac{1}{2}k_0 w_{0y}^2$ are Rayleigh ranges corresponding to the transverse scale widths, and $\alpha_0 = w_{0\xi}/w_{0x}$ is the spacetime eccentricity of the STOV.

For the STOV pulse of Eq. (1) launched from vacuum into a uniform slab of dispersive medium with dimensionless group velocity dispersion $\beta_2$ (see Appendix A), the expectation value of intrinsic electromagnetic tOAM in the medium *per incident photon* is [2] $\langle L_y^{EM} \rangle_{med} = \frac{1}{2} l \left( \alpha - \frac{\beta_2}{\alpha} \right)$. If the incident pulse eccentricity is $\alpha_0$, then in the medium $\alpha = n_g^{-1} \alpha_0$ due to the longitudinal spatial compression of the STOV upon entering the slab, where $n_g = c/v_g$ is the group index in the medium. As the tOAM per incident photon is $\langle L_y^{EM} \rangle_{inc} = \frac{1}{2} l \alpha_0$, this suggests that *the medium itself* has taken up intrinsic tOAM per incident photon of

$$\langle L_y^{med} \rangle = \frac{1}{2} l((\alpha_0 - \alpha) + \beta_2/\alpha) = \frac{1}{2} l(\alpha_0(1 - n_g^{-1}) + \beta_2/\alpha) , \tag{2}$$

assuming negligible reflection. We have identified this material-based tOAM structure as a STOV polariton [2]. Both terms in Eq. (2) are generated by pondermotive forces, as discussed below and in [16]. For a dispersionless medium, the first term in Eq. (2) becomes $\frac{1}{2} l \alpha_0 (1 - n^{-1})$, where $n = ck_0/\omega$ is the refractive index, and the second term vanishes. For a collisionless plasma of electron density $N_e$ with dielectric function $\varepsilon(\omega) = n^2(\omega) = 1 - \omega_p^2/\omega^2$, the dimensionless GVD is $\beta_2 = -\omega_p^2/\omega_0^2 = -N_e/N_{cr}$, where $\omega_p = (4\pi N_e e^2/m)^{1/2}$ is the plasma frequency, $N_{cr} = m\omega_0^2/4\pi e^2$ is the critical density, and $\omega_0$ is the central frequency. The group velocity is $v_g = c(1 - N_e/N_{cr})^{1/2}$.

The PIC simulation (described in Appendix B) computes the $\mathbf{E}$ and $\mathbf{H}$ fields in vacuum and in the plasma as well as the positions $\mathbf{r}_j$ and momenta $\mathbf{p}_j$ of all the particles in the plasma. The tOAM of the EM field and the plasma medium are, per incident photon,

$$L_y^{EM} = 2k_0 U_{inc}^{-1} \int d^2\mathbf{r}_\perp d\xi (\mathbf{r} \times (\mathbf{E} \times \mathbf{H}^*))_y \tag{3a}$$

$$\begin{aligned} L_y^{med} &= L_{y,elec} + L_{y,ion} \\ &= 8\pi c k_0 U_{inc}^{-1} \left[ \sum_j W_j (\mathbf{r}_{j,elec} \times \mathbf{p}_{j,elec})_y + \sum_j W_j (\mathbf{r}_{j,ion} \times \mathbf{p}_{j,ion})_y \right], \end{aligned} \tag{3b}$$

where $U_{inc} = \int d^2\mathbf{r}_\perp d\xi (|\mathbf{E}|^2 + |\mathbf{H}|^2)$ is evaluated at $z < 0$ before the pulse encounters the entrance interface, $W_j$ is the number of electrons or protons in a "macroparticle" (see Appendix B), the index $j$ ranges over all the electron and proton macroparticles, and the subscripts $elec$ and $ion$ refer to the electrons and ions (protons). Here and elsewhere in the paper we use Gaussian units. The centres of energy of both the STOV pulse in vacuum (symmetric in its energy density) and the composite field-particle disturbance in the medium propagate along $x = 0$ in the $\hat{\mathbf{z}}$ direction**.** This is because the direction of the total linear momentum, $\mathbf{P} = \mathbf{v}_{\mathrm{CE}}U/c^2$, is unchanged in the passage of the pulse from vacuum into the medium. Here, $U = \int d^3\mathbf{r}\, u(\mathbf{r},t) + \sum_i \gamma_i m_i c^2$ is the total energy of the fields plus particles, $\mathbf{v}_{\mathrm{CE}} = d\mathbf{r}_{\mathrm{CE}}/dt$, and $\mathbf{r}_{\mathrm{CE}} = U^{-1}[\int d^3\mathbf{r}\, \mathbf{r}\, u(\mathbf{r},t) + \sum_i \mathbf{r}_i \gamma_i m_i c^2]$ is the centre of energy, where $u(\mathbf{r},t)$ is the EM energy density, and $\mathbf{r}_i(t)$, $\gamma_i$ and $m_i$ are the particle positions, Lorentz factors, and rest masses. Therefore, choosing the origin $(x,z) = (0,0)$ for calculation of $L_y^{EM} + L_y^{med}$ from Eq (3) ensures that extrinsic tOAM vanishes [14] and that the computed tOAM is intrinsic. All tOAM discussed in this paper is intrinsic. An auxiliary computation inserting PIC results into the above expression for $\mathbf{r}_{\mathrm{CE}}$ verified that $x_{CE} = 0$ within numerical error.

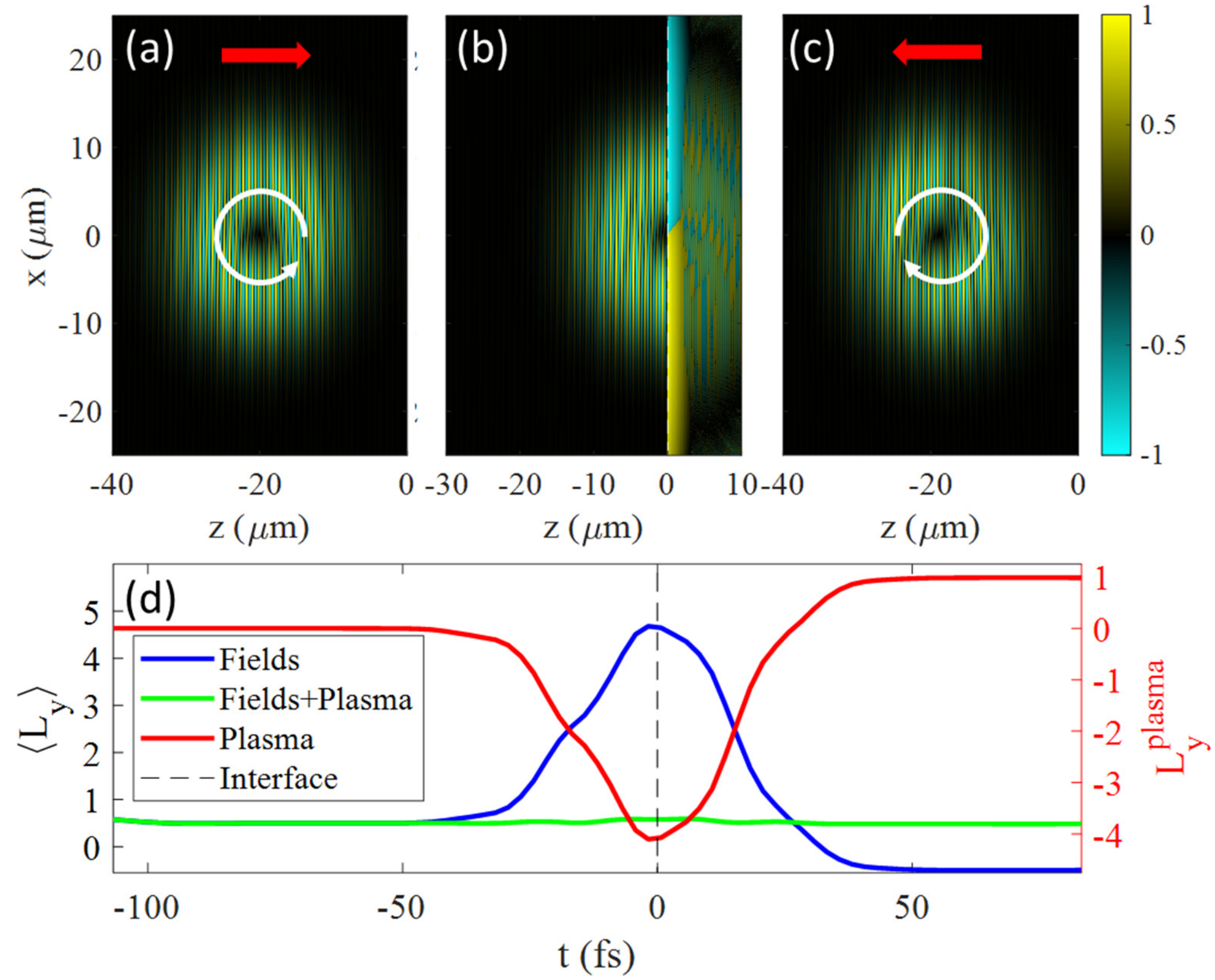

**Figure 1.** **(a)** Propagation of an $l = 1$, $\alpha_0 = 1$, $a_0 = 0.001$ STOV pulse toward an overcritical plasma slab ($N_e/N_{cr} = 2$) with front face at $z = 0$. The phase winding direction is superimposed. **(b)** STOV singularity has reached the interface, with half of the pulse reflecting and half still incident; the log of the evanescent field is plotted at $z > 0$. **(c)** STOV fully reflected from interface, with winding and tOAM flipped. **(d)** Time evolution of tOAM contributions through the interaction. Units are tOAM per incident photon. The red scale at the right applies to the red curve. The black scale at the left applies to the other curves. The dashed black line at $t = 0$ marks the arrival of the STOV singularity at the interface, corresponding to panel (b).

We first examine the normal incidence reflection of a $y$-polarized STOV pulse from an overcritical plasma slab with $N_e/N_{cr} = 2$, whose front face is at $z = 0$. This plasma density ensures 100% energy reflection over the full STOV bandwidth. Initially, we consider the linear case for small normalized vector potential, $a_0 = eA_0/mc^2 = 0.001$, where $A_0$ is the peak vector potential. We use $l = 1, \alpha_0 = 1$, $\lambda_0 = 800\, nm$ (for which $N_{cr} = 1.71 \times 10^{21}\ \mathrm{cm}^{-3}$), $w_{0\xi} = 10\ \mu m$ (FWHM pulsewidth $\tau \sim 55\ fs$), $w_{0x} = 9\ \mu m$ (so that $z_{0x} \sim 300 \mu m$), and $w_{0y}/w_{0x} \gg 1$ so that the field dependence on $y$ can be neglected. Launched from its waist at $z = -50\ \mu\mathrm{m}$, the STOV approaches the slab's front face in Fig. 1(a). In Fig. 1(b), the STOV singularity has reached the interface, with the leading part of the pulse reflecting while the back of the pulse is still incident. The fully reflected pulse is shown in Fig. 1(c), where the reflection boundary conditions ensured that the STOV fields, phase winding, and tOAM flip sign.

The time-dependent accounting of tOAM contributions is plotted in Fig. 1(d), where time $t$ tracks propagation of the STOV singularity in the lab frame. As the STOV approaches the interface (Fig. 1(a)), it has +½ unit of tOAM per photon (blue curve for $t < \sim -50$ fs), as predicted by our theory [2]. As interaction with the plasma begins, a transient positive spike in the EM tOAM develops, accompanied by an opposing spike associated with the plasma tOAM (red curve). The vertical dashed line marks the arrival of the STOV singularity at the interface at $t = 0$ (corresponding to Fig. 1(b)). After reflection is complete ($t > \sim 40\ fs$), the STOV departs in the $-\hat{\mathbf{z}}$ direction with −½ units of tOAM per photon. The plasma is left with twice the tOAM of the incident STOV, +1 unit per incident photon (red curve, $t > \sim 40$ fs), conserving tOAM. The sum intrinsic tOAM of the fields and particles per incident photon is plotted as the green curve in Fig. 1(d); it remains constant at +½ throughout. For an incident STOV with $l = -1$, the spikes flip and the sum tOAM remains constant at −½. As will be discussed later, the source of the counterbalancing interface spikes in tOAM for fields and particles is the spatiotemporal torque exerted by the light on the plasma and vice versa. Additionally, we note that these results confirm the half-integer values of intrinsic tOAM and the transmission of tOAM to material particles, contradicting the claims in [17,18].

Next, we examine the same weak STOV pulse normally incident on a thin ($d = 50\ \mu m$, $d/z_{0x} \ll 1$) subcritical density plasma slab. In Fig. 2(a) ($N_e/N_{cr} = 0.0025$), the STOV's tOAM per photon (blue curve) is +½ as it approaches the front interface at $z = 0$ and then spikes as it enters the plasma, with equal and opposite tOAM picked up by the plasma particles (red curve). These spikes flip for $l = -1$. As the STOV fully enters the plasma, the tOAM levels off to a steady value. We later compare simulation results in this steady propagation region with theory [2].

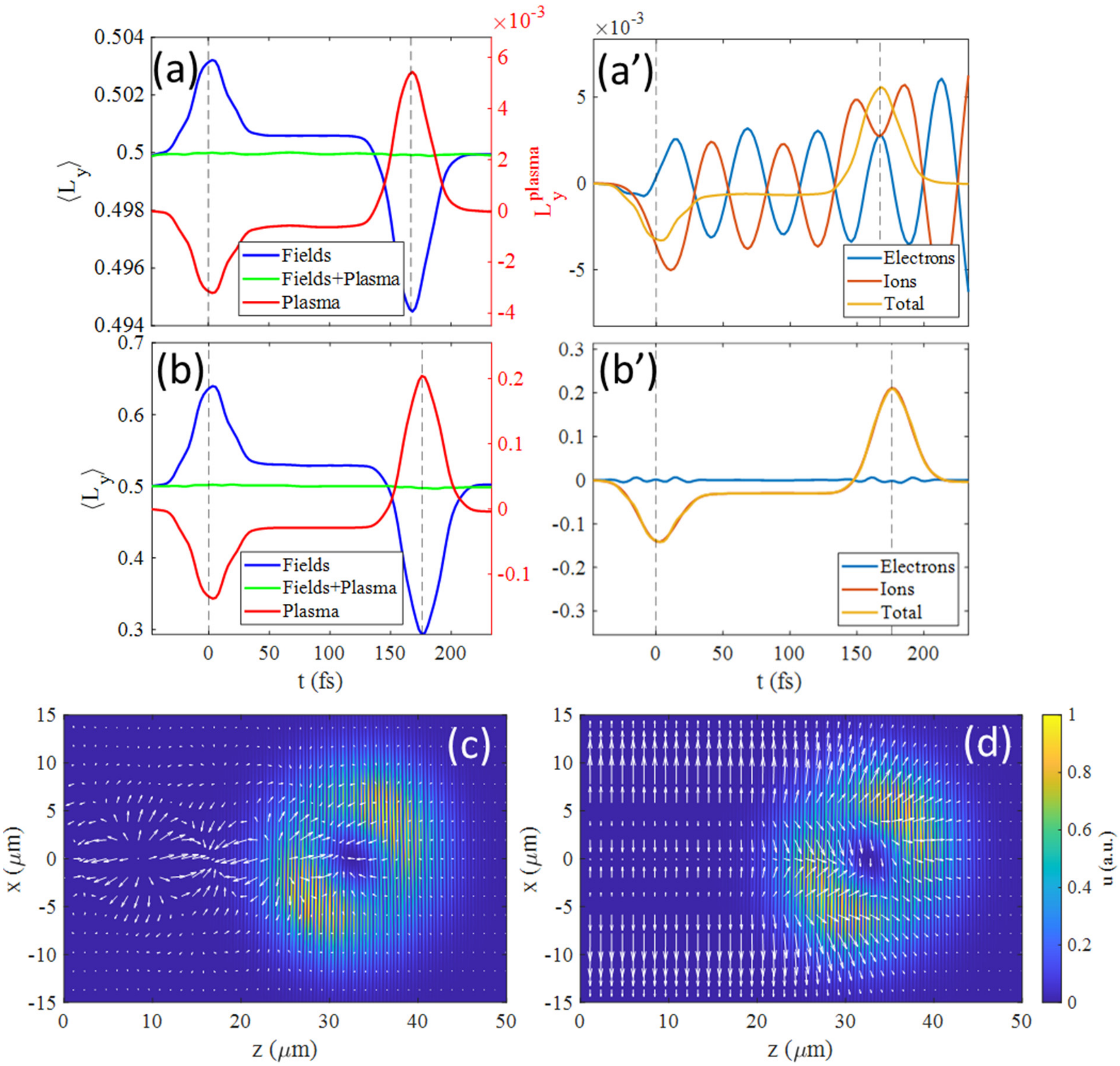

**Figure 2. (a)** Intrinsic tOAM of STOV pulse (linear case $a_0 = 0.001$), plasma, and their sum vs. propagation distance through a slab $d = 50\ \mu m$ thick with $N_e/N_{cr} = 0.0025$. Here $z_{0x} \cong 390\mu m$, so that $d/z_{0x} \ll 1$. The constant sum verifies conservation of intrinsic tOAM over the particles and fields. The vertical dashed lines denote the times when the STOV singularity passes through the slab entrance and exit faces. **(a')** Electron and ion tOAM, and their sum (corresponding to (a)). **(b)** same as (a) for $N_e/N_{cr} = 0.1$ . **(b')** Electron and ion tOAM, and their sum (corresponding to (b)). For conditions of (a), (a'): **(c)** electron linear momentum (white arrows) and **(d)** Electron + ion total linear momentum (white arrows), both in the lab frame and superimposed on the STOV intensity profile $u$. The arrows illustrate direction and relative amplitude, with each panel's arrow lengths separately normalized. Interface reflections are automatically accounted for in all of these simulations.

As the STOV passes through the right side of the slab, a negative (positive) spike in field (particle) tOAM develops. Upon exit, the STOV emerges with its original +½ tOAM per photon, as the particles have returned all tOAM to the field. Throughout the propagation, the sum tOAM (green curve) remains +½.

Results for the higher density case $N_e/N_{cr} = 0.1$ are plotted in Fig. 2(b). The form of the tOAM evolution is similar to Fig. 2(a), with counterbalancing positive and negative spikes of the field and particle contributions at the slab entrance and exit, with particle tOAM returned to the exiting pulse. However, because of the much higher plasma density, the EM and particle tOAM levels in the medium are commensurately higher.

The individual responses of the electrons and ions for Fig. 2(a) and 2(b) are plotted in Fig. 2(a') and 2(b'); the responses are strongly dependent on plasma density. In Fig. 2(a,a'), for $w_{0\xi} < \lambda_p = 2\pi c/\omega_p$, the tOAM of the electrons and ions oscillate $\pi$ out of phase, exchanging tOAM at the plasma frequency $\omega_p$. The tOAM oscillation amplitudes are similar because the electrostatic forces and the imparted momentum are similar for both species. After the pulse exits the slab, the oscillations in electron and ion tOAM continue, but they sum to zero. In Fig. 2(b,b'), where $w_{0\xi} \gg \lambda_p$, torque by the light on the electrons is immediately distributed between the electrons and ions according to their mass ratio; here the pulse is long enough for plasma electrostatic forces to pull the ions to the same velocity as electrons. The sum particle tOAM goes to zero as the pulse exits the slab. Figures 2(c) and 2(d) correspond to the conditions of Figs. 2(a,a′), and plot the lab frame electron momentum and total (electron plus ion) linear momentum evolution driven by the STOV pulse within the slab. These panels are discussed below in the context of Fig. 4.

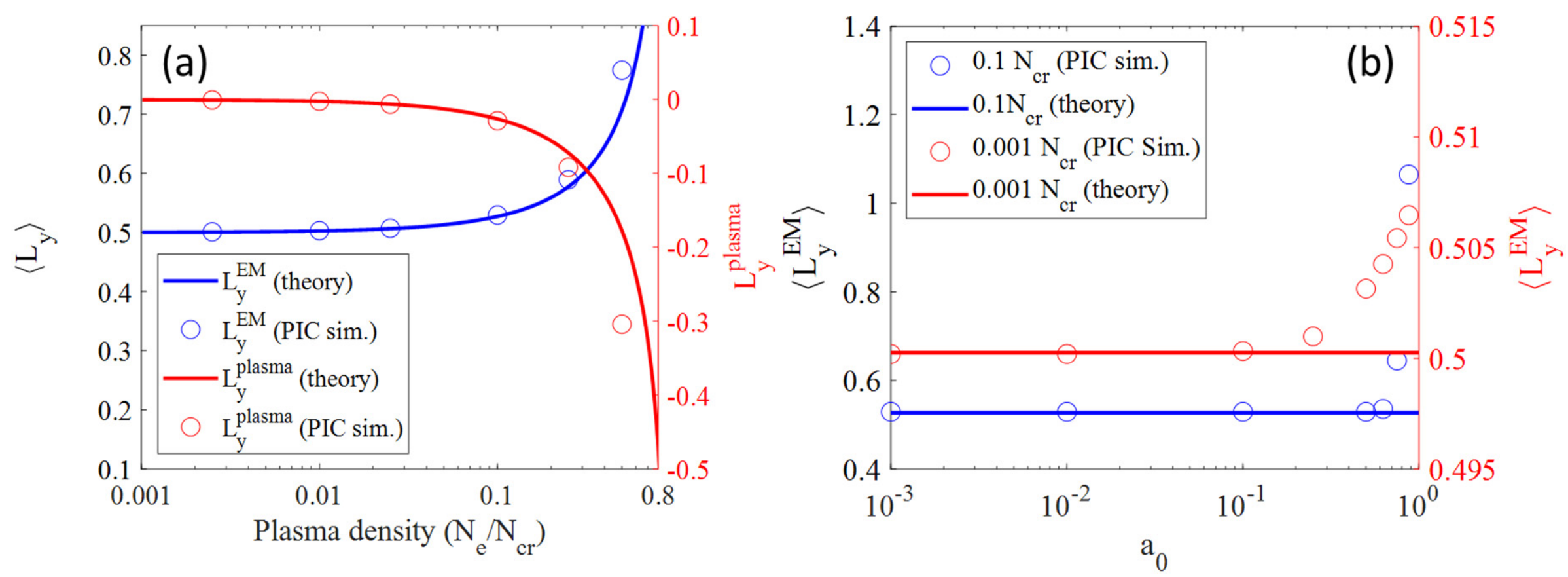

**Figure 3.** Comparison of tOAM linear theory [2] and PIC simulation results for varying plasma density and peak field. **(a)** Field tOAM $\langle L_y^{EM} \rangle$ and particle tOAM $L_y^{plasma}$ per incident photon from theory (solid curves) and PIC simulation (open circles) for $a_0 = 0.001$. Deviation from theory occurs only at near-critical density, where the redder portion of the incident STOV spectrum cannot penetrate the interface. **(b)** Comparison of linear theory (solid curves) and PIC simulation predictions of $\langle L_y^{EM} \rangle$ for two densities, for $a_0$ ranging from linear to relativistic strength ($a_0 \sim 1$).

We now consider the steady tOAM of the EM field and particles during propagation of the STOV in the bulk medium away from the interfaces. Plotted in Fig. 3(a) (blue curve) is the theory prediction for the EM tOAM in the medium $\langle L_y^{EM} \rangle = \frac{1}{2} l(\alpha - \beta_2/\alpha) = \frac{1}{2}[v_g/c + (c/v_g) N_e/N_{cr}]$ for $l = 1$ and $\alpha_0 = 1$. In all panels of Fig. 3, the theory points are adjusted to account for reflections at the interfaces. The PIC simulation (blue circles, from Eq. (3a)) for $a_0 = 0.001$ is in excellent agreement with the theory. For tOAM of the medium, the theory predicts $\langle L_y^{med} \rangle = \frac{1}{2}(1 - v_g/c - (c/v_g) N_e/N_{cr})$, plotted as $L_y^{plasma}$ (red curve) in Fig. 3(a). The PIC simulation results (using Eq. (3b)) are plotted as the open red circles. Agreement with the theory is excellent, *confirming both the existence of STOV polaritons and the value of their tOAM*. The PIC results deviate only near critical density, where red frequencies in the STOV bandwidth reflect more than blue from the entrance interface, affecting both particle and EM tOAM.

Up to this point, we have examined only linear propagation of STOVs in dispersive media. Figure 3(b) shows how the simulation results deviate from the linear theory as the field is increased to relativistic levels, $a_0 \to 1$. $\langle L_y^{EM} \rangle$ is plotted for two cases, $N_e/N_{cr} = 0.001$ and $N_e/N_{cr} = 0.1$. For the low-density case, theory matches the PIC simulation up to $a_0 \sim 0.2$, while for the higher density case it matches up to $a_0 \sim 0.7$. While relativistic self-focusing is weak for our thin slab ($d/z_{0x} \ll 1$), ponderomotive expulsion of electrons begins as $a_0 \to 1$. For the higher density case, expulsion is weaker owing to the stronger electrostatic restoring force on the electrons; hence the better fit of the linear theory for larger $a_0$. Simulation points deviating from the theory for high $a_0$ were computed as propagation averages; $\langle L_y^{EM} \rangle$ begins to fluctuate from excitation of large amplitude plasma waves, which collectively take up tOAM.

Left unaddressed thus far is the nature of the STOV polariton. We start with and then justify the statement that *it is ponderomotive forces at the medium interface and in the presence of material dispersion—originating from the magnetic Lorentz force—that drive the STOV polariton*. In the supplementary material [16] we show this for a general dielectric medium. Here we specify to the case of plasma.

To have OAM along $\hat{\mathbf{y}}$, there must be particle linear momentum in the spatiotemporal or $x\xi$ plane. The only driver for such motion in plasma is the magnetic Lorentz force density, $\mathbf{f} = N_e(-e)\mathbf{v} \times \mathbf{B}/c$, where $\mathbf{v}$ is electron velocity and $\mathbf{B}$ is the laser magnetic field. After cycle averaging, this becomes the ponderomotive force $\bar{\mathbf{f}} = -\nabla u_{pond}$, where $u_{pond} = \frac{1}{4} N_e mc^2 a^2$ is the ponderomotive energy density ($a = eA/mc^2$) [16]. Note that $\nabla$ rather than $\nabla_{ST}$ (see Appendix A) is used because the plasma is stationary in the lab frame. This force can also be derived using the divergence of the electromagnetic stress tensor [19] in the plasma dielectric medium.

Pondermotive forces generate the terms in Eq. (2) as follows: The first term, non-zero even for dispersionless media, originates from the net torque applied to the material from the imbalance across $\xi$ of $x$-directed EM linear momentum flow as the pulse passes through the interface. The material dispersion responsible for the second term promotes an imbalance across $x$ of $\xi$-directed EM linear momentum flow. All these momentum flow gradients present as ponderomotive forces. To understand the role of dispersion, Fig. 4(a) presents a Wigner plot (see Appendix C) of the incident STOV pulse. The $x - \tau$ projection is the familiar spacetime STOV intensity profile, $|E(x,\xi)|^2$, where $\xi = v_g \tau$. The $x - \omega$ projection is the spatiospectral profile $|\tilde{E}(x,\omega)|^2$, which exhibits a spatiospectral "tilt" as it enters the medium. Given that the plasma dielectric response depends on frequency, it is evident that the distribution of ponderomotive force density will be unbalanced along $x$, through the centre of energy of the pulse, resulting in a net torque. Plotted in Figs. 4(b) and 4(c) are maps of the $y$-component of torque density on the plasma, $T_y(x,\xi) = (\mathbf{r} \times \bar{\mathbf{f}})_y = (\xi\, \partial/\partial x - x\, \partial/\partial \xi) u_{pond}$, where $u_{pond}$ is taken from the PIC simulation. In Fig. 4(b), as the front part of the pulse enters the medium, but before the STOV singularity reaches the interface, the integral of the torque density is negative (hence the negative-going spike in particle tOAM in Figs. 2(a) and 2(b)). As the singularity moves past the interface and into the bulk, the torque density integral becomes positive, the spike declines, and the particle tOAM relaxes to its steady value (Fig. 4(c)). As seen in Figs. 2(a) and 2(b), as the STOV pulse exits the slab, the opposite effects occur. In Fig. 4(d) we compare the time derivative of the plasma tOAM of Fig. 2(b), $dL_y^{plasma}/dt$, and the integrated torque density $T(t) = \int dxd\xi\ \mathcal{T}_y(x,\xi;z)$, where $t = z/v_g$. The excellent agreement supports our picture of spatiotemporal torques in dispersive media driven by the magnetic Lorentz force (or cycle-averaged, the ponderomotive force). In fact, the robustness of the linear theory–even for near-relativistic $a_0$–is likely due to the leading order role of the magnetic Lorentz force in the electron dynamics.

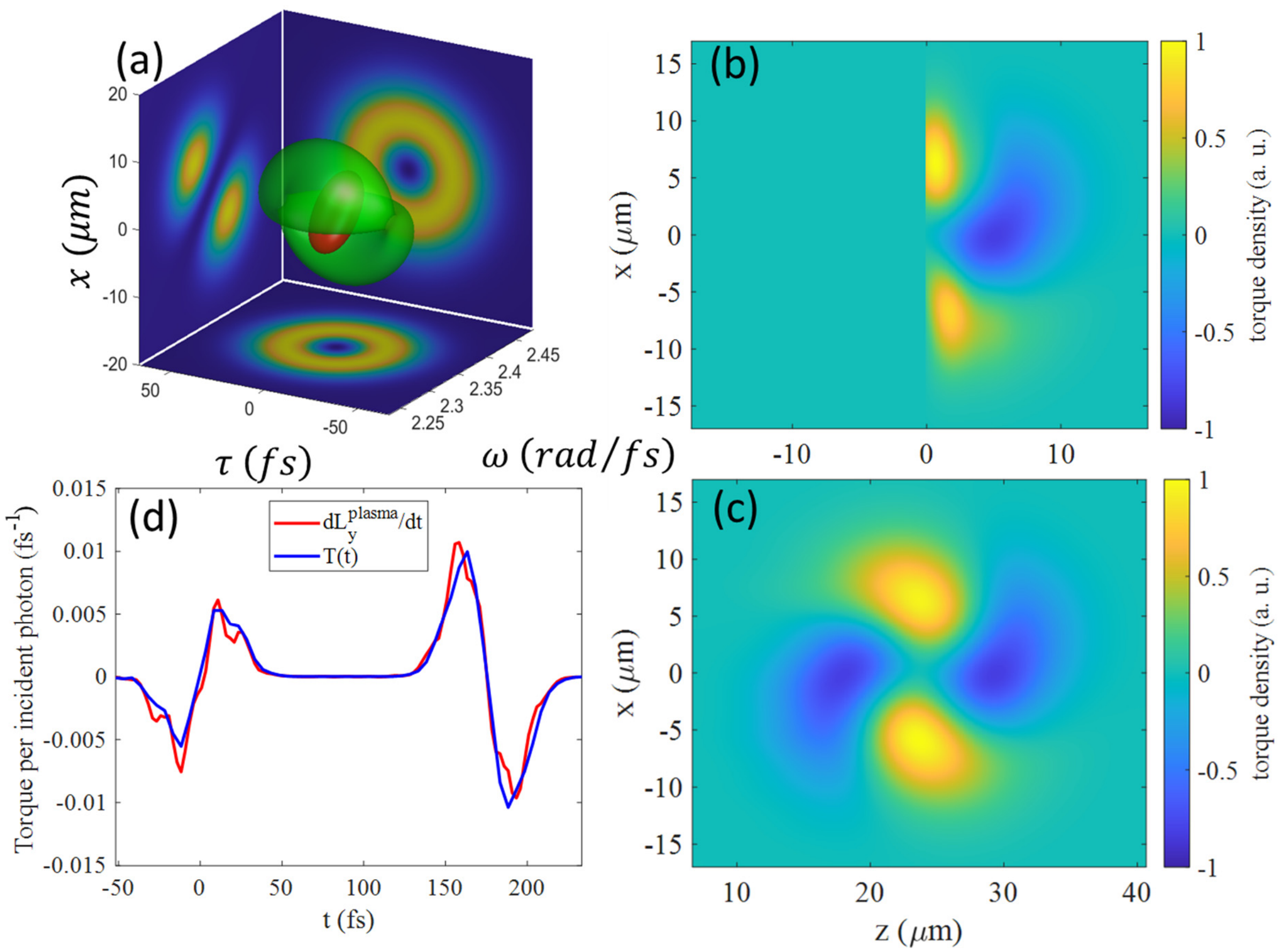

**Figure 4. (a)** Wigner plot of the STOV pulse before it enters the interface and its integral projections (see Appendix C). The central object (the Wigner intensity distribution) is the 20% intensity isosurface; green depicts where the field is positive and red depicts where it is negative. The familiar spacetime STOV is seen in the $x-\tau$ projection $|E(x,\xi)|^2$ (where $\xi = v_g\tau$), its spatiospectral representation $|\tilde{E}(x,\omega)|^2$ is the $x-\omega$ projection, and its spectral evolution is in the $\omega-\tau$ projection. **(b)** Torque density $\mathcal{T}_y(x,\xi)$ on electrons just before the STOV singularity reaches the front interface. **(c)** Torque density on electrons when the STOV is entirely inside the slab. The plasma density is $N_e/N_{cr} = 0.1$ and $a_0 = 0.001$. (c) and (d) are normalized since torque density scales linearly with pulse energy. **(d)** Plots of time derivative of the plasma tOAM of Fig. 2(b), $dL_y^{plasma}/dt$, and of integrated torque density $T(t) = \int dxd\xi\mathcal{T}_y(x,\xi;z)$, where here $t = z/v_g$.

Given this torque picture, we now reexamine Figs. 2(c) and 2(d) ($N_e/N_{cr} = 0.0025$ and $a_0 = 0.001$), where the STOV pulse is fully in the bulk. The oscillations in Fig. 2(c) consist of phased rotations of the local electron momentum vector, with the oscillations (at the plasma period $2\pi/\omega_p$) persisting well after the pulse (as also seen in Fig. 2(a′)). In Fig. 2(d), the local net (electron + ion) momentum vector ceases to rotate after the pulse passes, leaving in its wake transversely directed linear momenta corresponding to mass outflow and inflow, in a pattern governed by the time-integrated pondermotive force profile. In both Figs. 2(c) and 2(d), the linear momentum vectors grow and rotate in response to the onset and passage of the torque density profile $\mathcal{T}_y(x,\xi)$ (Figs. 4(b) and (c)). However, the integral of the induced tOAM density has a non-zero contribution only in the vicinity of the driving STOV pulse, which is where $\mathcal{T}_y(x,\xi)$ increases the medium tOAM from zero to the steady value of the propagating polariton. This is more clearly seen in Fig. 2(d), where inclusion of the ions ensures that the local total momentum stops rotating as the pulse passes by and $\mathcal{T}_y(x,\xi) \to 0$, revealing the instantaneous linear momentum structure of the STOV polariton superimposed on the STOV. Finally, we emphasize that these ideas are applicable to a general uniform dielectric medium, where the free or bound current density driven by the cycle-averaged magnetic Lorentz force (or the pondermotive force) leads to the medium tOAM per photon of $\langle L_y^{med} \rangle = \frac{1}{2} l(\alpha_0 - \alpha) + \frac{1}{2} l\, \beta_2/\alpha$ [16].

In summary, we have confirmed the existence and physical origin of a new quasiparticle in bulk dispersive media, the STOV polariton, and determined the value of its transverse orbital angular momentum (tOAM), verifying our theoretical predictions based on a given fixed dispersion [2]. Employing a first principles dynamical model for the medium response—Maxwell-Lorentz dynamics applied to a fully ionized hydrogen plasma—we find that the polariton and its tOAM are excited by torques imposed by the magnetic Lorentz force, even for weak fields. This is new type of laser-matter torque unanticipated in our prior work [14], and may play a role in future photonic technologies based on magneto-electric effects [20].

This work is supported by the Air Force Office of Scientific Research (FA9550-21-1-0405) and the US Department of Energy (DE-SC0024398 and DE-SC0024406).

## APPENDIX A: REVIEW OF TRANSVERSE OAM THEORY

To briefly review, the linear theory finds modal solutions to the spatiotemporal paraxial wave equation in dispersive media,

$$2ik_0\, \partial\mathbf{A}/\partial z = (-\nabla_\perp^2 + \beta_2\, \partial^2/\partial\xi^2)\mathbf{A} = H\mathbf{A} \ , \tag{A1}$$

where $\mathbf{A} = \mathbf{A}(\mathbf{r}_\perp, \xi; z)$ is the propagating field, $\mathbf{r}_\perp$ represents coordinates transverse to the propagation direction $\hat{\mathbf{z}}$, $\nabla_\perp^2$ is the transverse Laplacian, $H$ is the propagation operator, $k_0 = k(\omega_0)$ is the pulse central wavenumber in the medium, $\xi = v_g t - z$ is a space coordinate local to the pulse, $v_g = (\partial k/\partial\, \omega)_{\omega_0}^{-1}$ is the pulse group velocity, $\beta_2 = v_g^2 k_0 k_0''$ is the dimensionless group velocity dispersion (GVD), and the argument $z$ is separated by a semicolon it plays the role of a timelike running parameter.

The modal solutions derived in [2] are $A_{mpq}(x,y,\xi;z) = A_{mpq}^{(0)} u_m^x(x;z) u_p^y(y;z) u_q^\xi(\xi;z)$, where $u_m^x(x;z) = C_m w_x^{-1/2}(z) H_m\left(\frac{\sqrt{2}\,x}{w_x(z)}\right) e^{-x^2/w_x{}^2(z)} e^{ik_0 x^2/2R_x(z)} e^{-i(m+1/2)\psi_x(z)}$ and $u_p^y(y;z)$ (with similar form) are standard Hermite-Gaussian functions with $C_m = (2/\pi)^{1/4}(2^m m!)^{-1/2}$, $H_m$ is a Hermite polynomial of order $m$, $w_x(z) = w_{0x}(1 + (z/z_{0x})^2)^{1/2}$, $R_x(z) = z(1 + (z_{0x}/z)^2)$ and $\psi_x(z) = \arctan(z/z_{0x})$. Meanwhile, $u_q^\xi(\xi;z) =$ $C_q w_\xi^{-1/2}(z) H_q\left(\frac{\sqrt{2}\,\xi}{w_\xi(z)}\right) e^{-\xi^2/w_\xi^2(z)} e^{-ik_0\xi^2/2\beta_2 R_\xi(z)} e^{i(q+1/2)\psi_\xi(z)}$, where $C_q = (2/\pi)^{1/4}(2^q q!)^{-1/2}$, $H_q$ is a Hermite polynomial of order $q$, $w_\xi(z) = w_{0\xi}(1+(z/z_{0\xi})^2)^{1/2}$, $R_\xi(z) = z(1+(z_{0\xi}/z)^2)$, $\psi_\xi(z) = \mathrm{sgn}(\beta_2)\arctan(z/z_{0\xi})$, and $z_{0\xi} = k_0 w_{0\xi}^2/2|\beta_2|$.

Any spatiotemporal field can be constructed as a linear combination of these spatiotemporal Hermite Gaussian (STHG) modes $A_{mpq}$. Azimuthal STOV modes of winding number $l$ (such as in Eq. (1)) can be represented in the STHG basis as [21]

$$\mathrm{STOV}_{0l}(x,y,\xi;z) = \sum_j^{|l|} C_j\, A_{j,0,|l|-j}(x,y,\xi;z), \tag{A2}$$

where $C_j = (|l|/(2j)!\,(|l|-j)!)\left([(|l|-2j)!\,(2j)!/2^{|l|}|l|!]\right)^{1/2}$. Radial STOV modes can be constructed in a similar manner.

In ref. [2], conservation of EM energy density flux leads to identification of the spacetime linear momentum operator $\hat{\mathbf{p}}_{ST}$ and the tOAM operator $L_y$:

$$\hat{\mathbf{p}}_{ST} = -i\nabla_{ST} = -i(\nabla_\perp - \hat{\boldsymbol{\xi}}\beta_2\,\partial/\partial\xi), \tag{A3a}$$

$$L_y = (\mathbf{r}\times\hat{\mathbf{p}}_{ST})_y = -i\left(\xi\frac{\partial}{\partial x} + \beta_2 x\frac{\partial}{\partial\xi}\right), \tag{A3b}$$

Furthermore, because the operators $H$ and $L_y$ commute, $[H, L_y] = 0$, the expectation value $\langle L_y\rangle = \langle A|L_y|A\rangle = \int d^2\mathbf{r}_\perp d\xi\, A^* L_y A$ of $L_y$ is conserved with propagation: $d/dz\,\langle L_y\rangle = i(2k_0)^{-1}\langle[H, L_y]\rangle = 0$. If $\langle L_y\rangle$ is calculated with respect to the pulse center of energy (or energy centroid), it is the *intrinsic* tOAM [14], a property of photons. In this paper, we consider only intrinsic tOAM.

## APPENDIX B: PARTICLE-IN-CELL SIMULATIONS

Our simulations were performed using the PIC code EPOCH [22], directly solving the Maxwell-Lorentz system of equations for the EM field and for the plasma electrons and ions (protons). The dimensions of the PIC simulation grid were $100\ \mu\mathrm{m}\ (z) \times 60\ \mu\mathrm{m}\ (x)$, with $4096 \times 1536$ grid points. The plasma initially contained a uniform distribution of 20 macroparticles per cell of each species (protons and electrons), with each macroparticle composed of $W_j = 8.15\times10^7$ to $1.63\times10^{11}$ protons or electrons depending on the defined density. The device of "macroparticles" is used to achieve realistic particle densities while greatly reducing the computational load. Crucially in our simulations, ions were allowed to move. While the protons' displacement is small compared to that of electrons owing to the high mass ratio, the laser field imparts approximately the same linear and angular momentum to both species; the proton response cannot be neglected.

## APPENDIX C: WIGNER FIELD DISTRIBUTION

The Wigner Field Distribution (WFD) gives the spectral composition of a signal as function of time [23].

$$\mathrm{WFD}(t,\omega) = \int_{-\infty}^{\infty} d\tau\, f\left(t+\frac{\tau}{2}\right) f^*\left(t-\frac{\tau}{2}\right) e^{-i\omega\tau}\,. \tag{A4}$$

It is straightforward to extend this to an electric field with spatial dependence to obtain the WFD as a function of $\{x, y, \xi, \omega\}$ :

$$\mathrm{WFD}(x,y,\xi,\omega) = \frac{1}{v_g}\int_{-\infty}^{\infty} d\xi'\, f\left(x,y,\xi+\frac{\xi'}{2}\right) f^*\left(x,y,\xi-\frac{\xi'}{2}\right) e^{-i\omega\xi'/v_g}\,. \tag{A5}$$

# Supplementary Material: Spatiotemporal optical vortex (STOV) polariton

M. S. Le, S.W. Hancock, N. Tripathi, and H.M. Milchberg

*Institute for Research in Electronics and Applied Physics and Dept. of Physics, University of Maryland, College Park, Maryland 20742, USA*

## I. Physical origin of the $\frac{1}{2}l(\alpha_0 - \alpha)$ and $l\frac{\beta_2}{2\alpha}$ terms in the material tOAM

Following [1,2], we write for the perpendicular and parallel components of the electromagnetic energy density flux inside a weakly dispersive medium, with respect to a moving window parametrized by $\xi = v_g t - z$:

$$\mathbf{j}_\perp = -i\frac{1}{2k_0}[A^*\nabla_\perp A - A\nabla_\perp A^*] \quad \text{(S1a)}$$

$$\mathrm{j}_\parallel = i\frac{\beta_2}{2k_0}\left[A^*\frac{\partial}{\partial\xi}A - A\frac{\partial}{\partial\xi}A^*\right]. \quad \text{(S1b)}$$

Here $A$ is the slowly varying envelope of the $y$-polarized field $\mathbf{A} = \hat{\mathbf{y}}A(x, y, \xi)\exp(ik_0 z - i\omega_0 t)$, $\beta_2 = v_g^2 k_0 k_0''$ is the dimensionless group velocity dispersion, $v_g = (\partial k(\omega)/\partial\omega)_{\omega_0}^{-1}$ is the group velocity, $k_0'' = (\partial^2 k(\omega)/\partial\omega^2)_{\omega_0}$, $k^2(\omega) = \omega^2\varepsilon(\omega)/c^2$ for $\mu = 1$, $k_0 = k(\omega_0) = nk_{vac}$ in the medium, and $k_{vac} = \omega_0/c$ in vacuum. Subscripts $\perp$ and $\parallel$ refer to components perpendicular and parallel to the pulse propagation direction. The relation between **j** and the Poynting flux **S** (in the moving window) is:

$$\mathbf{S}_\perp = \mathbf{j}_\perp, \quad \mathrm{S}_\parallel = -j_\parallel \quad \text{(S2)}$$

We now consider the difference in the cycle-averaged linear momentum density (in Gaussian units) between vacuum and inside the material,

$$\Delta\mathbf{p} = \frac{1}{8\pi c}(\mathbf{S}^{vac} - \mathbf{S}) = \frac{1}{8\pi c}[(\mathbf{j}_\perp^{vac} - \mathbf{j}_\perp) + (-j_\parallel^{vac} + j_\parallel)\hat{\boldsymbol{\xi}}] \quad \text{(S3)}$$

We identify this difference as the linear momentum density $\mathbf{p}_m = \Delta\mathbf{p}$ invested in the medium.

### 1. Origin of the $\frac{1}{2}l(\alpha_0 - \alpha)$ term

This term originates from the "$\perp$" term in Eq. (S3). As the STOV enters the medium from vacuum, its asymmetry ratio changes from $\alpha_0$ to $\alpha$ owing to axial spatial compression along $\xi$. This change in spatial distribution of the field in the $(x, \xi)$ plane is one source of change in electromagnetic tOAM which must be taken up by the medium via a torque. The torque occurs from the imbalance of transverse ($x$) linear momentum density across the interface, which presents as an axial gradient in the ponderomotive force, as seen below.

The transverse component of the medium's linear momentum density is $\Delta\mathbf{p}_\perp = \mathbf{p}_{m\perp} = p_{mx}\hat{\mathbf{x}}$, where

$$p_{mx} = \frac{1}{8\pi c}(j_x^{vac} - j_x) =$$
$$-i\frac{1}{16\pi c k_{vac}}\left(\left[A_{vac}^*\frac{\partial}{\partial x}A_{vac} - A_{vac}\frac{\partial}{\partial x}A_{vac}^*\right] - \frac{k_{vac}}{k_0}\left[A^*\frac{\partial}{\partial x}A - A\frac{\partial}{\partial x}A^*\right]\right) \tag{S4}$$

Here, $A_{vac}$ is of the form of Eq. (1) of the main paper. But inside the medium $A(x,\xi;z=0) = A^{(0)}\sqrt{\frac{c/n}{v_g}}\left(\frac{c\xi}{v_g w_{0\xi}} + \text{sgn}(l)\ \frac{ix}{w_{0x}}\right)^{|l|} e^{-(x^2/w_{0x}^2+y^2/w_{0y}^2)}e^{-(c\xi/v_g w_{0\xi})^2}$ owing to the axial spatial compression.

The contribution of $p_{mx}$ to the y-component of tOAM density is

$$(\mathbf{r}\times\mathbf{p}_{m\perp})_y = \ \xi p_{mx}\ , \tag{S5}$$

from which we compute the medium's tOAM per incident photon for a STOV of charge $l$ and spatiotemporal asymmetry $\alpha_0$,

$$\langle L_y^{med1}\rangle = \omega_0\frac{\int d^3\mathbf{r}\,(\mathbf{r}\times\mathbf{p}_{m\perp})_y}{\int d^3\mathbf{r}\,u_{vac}} = \frac{l}{2}\alpha_0\left(1-n_g^{-1}\right) = \frac{l}{2}(\alpha_0\ -\ \alpha)\,. \tag{S6}$$

Here $n_g = c/v_g$ is the group index, $\alpha = \alpha_0/n_g$, the vacuum energy density is $u_{vac} = |A|^2/8\pi$, and the superscript $med1$ refers to the first term in Eq. (2) of the main paper. The role of the ponderomotive force (also see Sec II below) can be inferred by using $A = ae^{i\Phi}$, where $a$ and $\Phi$ are real. Then $p_{mx} \propto -i(A^*\frac{\partial}{\partial x}A - A\frac{\partial}{\partial x}A^*) = 2a^2\frac{\partial\Phi}{\partial x}$, and the force density associated with this term is ponderomotive: $f_x = \frac{\partial}{\partial t}p_{mx} \propto \frac{\partial}{\partial\xi}p_{mx} \propto l\frac{\partial}{\partial\xi}(a^2/\xi)$, with $\frac{\partial\Phi}{\partial x} = l/\xi$ for a STOV of charge $l$. The integral of the torque density term $\xi f_x$ has a non-zero contribution to the tOAM *only* as the STOV passes through the interface. Once the STOV is fully in the bulk, the integral goes to zero.

## 2. Origin of the $l\frac{\beta_2}{2\alpha}$ term

This term originates from the "$\parallel$" term in Eq. (S3). The parallel component of the medium's linear momentum density is $\Delta\mathbf{p}_\parallel = \mathbf{p}_{m\parallel} = p_{m\xi}\hat{\boldsymbol{\xi}}$ . Noting that $j_\parallel^{vac} = 0$, this term's contribution to the $y$-component of angular momentum density of the medium is

$$(\mathbf{r}\times\mathbf{p}_{m\parallel})_y = xp_{m\xi} = -i\frac{x\beta_2}{16\pi c k_0}\left[A^*\frac{\partial}{\partial\xi}A - \ A\frac{\partial}{\partial\xi}A^*\right]. \tag{S7}$$

Then, for an incident STOV of charge $l$ and spatiotemporal asymmetry $\alpha_0$ [1], the transverse OAM (tOAM) per incident photon is

$$\langle L_y^{med2}\rangle = \omega_0\frac{\int d^3\mathbf{r}\,(\mathbf{r}\times\mathbf{p}_{m\parallel})_y}{\int d^3\mathbf{r}\,u_{vac}} = l\frac{\beta_2}{2\alpha}\ . \tag{S8}$$

where $\alpha = \alpha_0/n_g$. Just as for the $\frac{1}{2}l(\alpha_0\ -\ \alpha)$ term, the space- and time-integrated torque density that leads to Eq. (S8) contributes only as the STOV pulse passes through the interface. We note that the numerator integrals in Eqs. (S6) and (S8) use the centre of energy as the origin, ensuring that $\langle L_y^{med}\rangle$ is intrinsic tOAM [3].

## II. Role of the magnetic Lorentz force

The driven current density in a general dielectric medium is $\mathbf{j}_m = \partial\mathbf{P}/\partial t$, where **P** is the medium's polarization. For time harmonic fields, $\tilde{\mathbf{j}}_m = -i\omega\widetilde{\mathbf{P}}$ and for a linear medium response, $\widetilde{\mathbf{P}} = \chi(\omega)\tilde{\mathbf{E}}$, where $\chi$ is the linear susceptibility of the medium. The magnetic force density on the material is then $\mathbf{f} = \frac{1}{2c}\,(\tilde{\mathbf{j}}_m \times \widetilde{\mathbf{B}}^* + \tilde{\mathbf{j}}_m^* \times \widetilde{\mathbf{B}})$. Using Maxwell's equation $\widetilde{\mathbf{B}} = \frac{-ic}{\omega}\nabla \times \tilde{\mathbf{E}}$, we get

$$f_i = \frac{1}{2}\left[\left(\chi E_j \frac{\partial}{\partial x_i} E_j^* + \chi^* E_j^* \frac{\partial}{\partial x_i} E_j\right) - \left(\chi E_j \frac{\partial}{\partial x_j} E_i^* + \chi^* E_j^* \frac{\partial}{\partial x_j} E_i\right)\right] , \tag{S9}$$

using summation over repeated indices ($j = 1,2,3$) and removing the tilde from the fields for convenience. The full vector form of the second term in brackets is $\chi(\mathbf{E}\cdot\nabla)\mathbf{E}^* + \chi^*(\mathbf{E}^*\cdot\nabla)\mathbf{E}$. This term is comparatively negligible for paraxial fields; this is the case for our situation. For a non-absorbing medium, $\chi$ is real and Eq. (S9) then becomes, after cycle averaging,

$$\bar{\mathbf{f}} = \frac{1}{4}\chi\nabla|\mathbf{E}|^2 = -\nabla u_{pond} , \tag{S10}$$

explicitly showing that the origin of the force density on the material is the magnetic Lorentz force, expressed as the negative gradient of the ponderomotive energy density $u_{pond} = -\frac{1}{4}\chi|\mathbf{E}|^2$.

For a collisionless (non-absorbing) plasma, the dielectric function is $\varepsilon = 1 + 4\pi\chi = 1 - \omega_p^2/\omega^2$, so that $\chi = -\,\omega_p^2/4\pi\omega^2$ and $u_{pond} = \frac{1}{16\pi}\frac{\omega_p^2}{\omega^2}|\mathbf{E}|^2 = N_e e^2|\mathbf{E}|^2/4m\omega^2$. In terms of the vector potential $A$,

$$\bar{\mathbf{f}} = -\frac{\omega_p^2}{16\pi c^2}\nabla|A|^2 = -\frac{1}{4}N_e mc^2\nabla a^2 , \tag{S11}$$

as presented in the main paper, where $a = e|A|/mc^2$ is the normalized vector potential.

Referenced to the lab origin, the torque density is $\mathcal{T}_y = z\bar{f}_x - x\bar{f}_z$ , where the overbar denotes cycle average. Employing the coordinate $\xi = v_g t - z$ gives $\mathcal{T}_y = (\xi\,\partial/\partial x - x\,\partial/\partial\xi)u_{pond}$, as presented in the paper.

---